\documentclass[twocolumn,amsmath,resetfootnote]{aastex702}

\usepackage{graphicx} 
\usepackage{booktabs}
\usepackage{tabularx}
\usepackage{array}

\definecolor{jaredpurple}{RGB}{93, 63, 211}
\definecolor{annablue}{RGB}{99, 162, 239}

\usepackage{ulem}
\definecolor{meridithgreen}{RGB}{0, 140, 0}
\long\def\mj#1{\textcolor{meridithgreen}{#1}}

\definecolor{lazgreen}{RGB}{93, 217, 128}

\definecolor{loganred}{RGB}{206, 50, 50}

\definecolor{samaqua}{RGB}{6, 145, 131}

\newcommand{\bbud}{\alpha\;\mathrm{Ori~B}}

\newcommand{\appropto}{\mathrel{\vcenter{
		\offinterlineskip\halign{\hfil$##$\cr
	\propto\cr\noalign{\kern2pt}\sim\cr\noalign{\kern-2pt}}}}}

\begin{document}

\title{Disco in the Dust: Reflected light at the bow shock around Betelgeuse's companion explains its observed luminosity}
\shorttitle{Betelgeuse Brightens Buddy's Bow-Shock}

\author[0000-0003-1012-3031]{Jared~A.~Goldberg}
\altaffiliation{NASA Hubble Fellow}
\affiliation{Columbia Astrophysics Laboratory, Columbia University, New York, NY, USA}
\affiliation{Department of Physics \& Astronomy, Michigan State University, MI, USA}
\email{goldstar@msu.edu}

\author[0000-0002-8717-127X]{Meridith Joyce} 
\affiliation{School of Physics and Astronomy, Rochester Institute of Technology, 1 Lomb Memorial Drive, Rochester, NY 14623, USA} 
\email{mxjsps@rit.edu}

\author[0000-0002-7296-6547]{Anna J. G. O'Grady}
\altaffiliation{McWilliams Fellow}
\affiliation{McWilliams Center for Cosmology and Astrophysics, Department of Physics, Carnegie Mellon University, Pittsburgh, PA, USA}
\email{aogrady@andrew.cmu.edu}

\author[0009-0000-9527-5477]{Sam~D.~Barber} 
\affiliation{Department of Physics \& Astronomy, University of Wyoming, 1000 E University Ave, Laramie, WY 02071, USA} 
\email{sbarber5@uwyo.edu}

\author[0000-0002-7174-8273]{Logan~J.~Prust} \affiliation{Center for Computational Astrophysics, Flatiron Institute, 162 Fifth Avenue, New York, NY 10010, USA} 
\email{lprust@flatironinstitute.org}

\author[0000-0001-9066-0552]{Mitchell~Dennis} 
\affiliation{Institute for Astronomy, University of Hawai`i at M\=anoa, 2680 Woodlawn Drive, Honolulu, HI 96822, USA}
\affiliation{School of Physics and Astronomy, Rochester Institute of Technology, 1 Lomb Memorial Drive, Rochester, NY 14623, USA}
\email{mtde226@hawaii.edu}

\author[0000-0002-8159-1599]{László~Molnár}
\affiliation{Konkoly Observatory, HUN-REN CSFK, Konkoly-Thege Mikl\'os \'ut 15-17, H-1121, Budapest, Hungary}
\affiliation{CSFK, MTA Centre of Excellence, Konkoly-Thege Mikl\'os \'ut 15-17, H-1121, Budapest, Hungary}
\affiliation{E\"otv\"os Lor\'and University, Institute of Physics and Astronomy, P\'azm\'any P\'eter s\'et\'any 1/A, H-1117, Budapest, Hungary}
\email{molnar.laszlo@csfk.org}

\author[0000-0002-8878-3315]{Christian~I.~Johnson}
\affiliation{Space Telescope Science Institute
3700 San Martin Drive
Baltimore, MD 21218, USA}
\email{}

\author[0000-0002-0479-7235]{Wolfgang~E.~Kerzendorf}
\affil{Department of Physics and Astronomy, Michigan State University, East Lansing, MI 48824, USA}
\affil{Department of Computational Mathematics, Science, and Engineering, Michigan State University, East Lansing, MI 48824, USA}
\email{wkerzend@msu.edu}


\begin{abstract}
Recent detections of $\alpha$~Orionis~B, the putative companion to Betelgeuse,
have been reported using multiple instruments and techniques. 
These include, most recently, a $>6\sigma$ detection using VLT/SPHERE reported by \citet{Montarges2026}. The authors infer a bright companion ($\sim10^{-3}\times$ Betelgeuse's luminosity) with a mass of $2-3\,M_\odot$ and $T_\text{eff}\approx10-12{,}000$\,K, loosely consistent with the uppermost bounds of \citet{Howell2025}'s recent mass estimate from speckle-imaging. 
However, it is in tension by a factor of two with the reported mass-exclusion limits via non-detection from a recent HST far-UV campaign \citep{Goldberg2025}. 
While mass identification from isochrone-fitting (conducted in all cases) is prone to uncertainty, the
restrictive HST upper limit on FUV flux 
precludes the hot, blue emission of a $\gtrsim2\,M_\odot$ main-sequence star. 
This discrepancy is reconciled by the following hypothesis: 
these detections observe the reflection of Betelgeuse's own luminosity scattered off the companion's bow shock and wake as it traverses the dusty circumstellar medium.
To evaluate the feasibility of this scenario, we draw from custom, filter-specific stellar models computed with MESA, as well as \texttt{Athena++} simulations of the companion's bow shock in idealized conditions. Importantly, the expansive bow shock from a stellar-mass object necessarily subtends a sizable portion of Betelgeuse's outgoing flux. 
We find that optical luminosity ratios of $\sim10^{-3}-10^{-4}$ are straightforward to resolve within reasonable assumptions about the companion mass and circumstellar dust, and disfavor a hot $3\,M_\odot$ companion. We thereby reconcile competing mass hypotheses across observational campaigns through proper attribution of the observed brightness, motivating
future multi-wavelength campaigns to further characterize this enigmatic system. 
\end{abstract}


\section{Introduction}\label{sec:intro}

Betelgeuse ($\alpha$~Orionis~A) is the closest red supergiant (RSG) to the Sun, and the $\sim$10th brightest star in our sky. Despite being studied for millennia, Betelgeuse's `Great Dimming' event in late 2019-early 2020  precipitated renewed interest in the star's variable nature, as it experienced its largest drop in apparent brightness in recorded memory \citep[][]{Guinan2019,Harper2020,Dharmawardena2020,Dupree2020,Dupree2022,Montarges2021,Taniguchi2022,Cannon2023,MacLeod2023,Jadlovsky2024}. 

Betelgeuse's cycles of variability include both periodic and stochastic signals in both its lightcurve and radial velocity (RV) timeseries. The stochastic component can be understood as arising from large-scale convection in its giant ($\approx800\,R_\odot$) outer H-rich envelope \citep[see, e.g.,][]{Goldberg2022a,Chiavassa2011,Chiavassa2024,JzMa2025,JzMa2026}. The periodic components include a $\approx36$\,yr period, often attributed to rotation \citep[e.g.][]{Wheeler2017,Kervella2018,Wheeler2023}, and $185$ and $\approx420$ day periods attributed to radial oscillations \mj{\citep{Joyce2020, Jadlovsky2023}}. Betelgeuse additionally exhibits a peculiar $\sim$2100-day `Long Secondary Period' (LSP) which defies simple pulsation models \citep[see extensive discussion by][and references therein]{Goldberg2024}. 

LSPs are common among cool, luminous evolved stars, seen in $\sim1/3$ of pulsating red giants, Asymptotic Giant Branch (AGB) stars, and RSGs \citep{Wood-2004,Soszynski-2021,Kiss-2006}. These signals are `long' in the sense that they are $\sim$a few$-10\times$ longer than the stars' fundamental radial pulsation modes (FMs; $\sim420$\,d in the case of Betelgeuse). 
The FM entails the longest-period oscillation expected to be present in these giant, convective stars. 

Several hypotheses for the driving mechanism of the LSP phenomenon have been proposed in the general case \citep{Wood-2004,Saio-2015,Takayama-2015,Soszynski-2021,Decin2025}. 
A leading hypothesis for the general LSP phenomenon is the presence of a binary companion\footnote{a.k.a. `buddy' or `little friend'} in orbit within the star's dusty surroundings \citet{Wood-1999,Wood-2004,Soszysnki-2007,Soszynski-2021}.\footnote{This hypothesis has been the subject of recent controversy in \textit{Gaia} astrometry, see \citet{Shariat-2026} and rebuttal by \citet{Iwanek-2026}.} 
In such a scenario, the period corresponds to the companion's orbit. However, when considering lightcurve and RV observations together, it is evident that the bright LSP phase tends to correspond to the companion's transit, and the dim phase to the companion's occultation \citep{Goldberg2024}. 
This defies a simple eclipsing dust-shrouded companion model, where transit should correspond to the dim phase. 
As such, recent proposed mechanisms for the LSP tend to instead invoke interaction between the companion and its environment --- for example, an  eccentric orbit passing through a surrounding dust shell \citep{Decin2025}, an expanding wake/long tidal tail \citep{MacLeod2025,Dupree2026,Matthews2026}, 
and generic `sculpting' of the circumstellar medium \citep[][not mutually incompatible with the other models mentioned]{Goldberg2024}. 
Although a companion's presence and orbital timescale can explain many aspects of the LSP phenomenon, the exact mechanism causing the photometric brightening and dimming remains an open puzzle.

In the case of Betelgeuse, a binary scenario was argued to be the only viable existing hypothesis for the cause of its $\sim$2100~day LSP \citep{Goldberg2024}. As such, \citet{Goldberg2024} and \citet{MacLeod2025} independently posited the presence of a binary companion, $\bbud$. Both works find companion masses between $\approx0.5-2M_\odot$ from the amplitude of the RV modulation, with an orbital separation $\approx2.5\times$ the radius of Betelgeuse ($D\approx{}2.5\,R_B$). Those results were presented in late 2024, shortly before the companion's predicted receding quadrature and maximum elongation from Betelgeuse in late November/early December. 
Since then, several studies have attempted to observe Betelgeuse's companion near maximum elongation, when the orbital configuration is most conducive separating the two sources on the sky with a predicted separation of $\approx53$\,mas. 
Non-detections were quickly reported in both X-ray \citep{OGrady2025} and far-UV \citep{Goldberg2025}, ruling out compact objects and bright accreting protostars $\gtrsim1.5\,M_\odot$.

The first `probable' detection of the companion was reported by \citet{Howell2025} from speckle imaging, finding a bright signal with low intrinsic significance ($\approx1.5\sigma$) 
but very close to the companion's predicted location as inferred from Betelgeuse's RV \citep{Goldberg2024, MacLeod2025} and posited rotational axis \citep{Kervella2018}. 
\citet{Howell2025} infer a mass of 1.4--2\,$M_\odot$, proposing the name \textit{Siwarha}, now recognized by the International Astronomical Union. 
Subsequently, \citet{Dupree2026} reported evidence of a `wake' coming from the companion traveling through Betelgeuse's chromosphere and dusty circumstellar wind, from periodic signals in optical and UV spectroscopy concurrent with the LSP. 
Finally, most recently, \citet{Montarges2026} report a $6.1\sigma$ detection with interferometric imaging using the VLT/SPHERE. Fitting main-sequence stellar spectra to the observed flux, they infer a bright young MS companion of $2.6-3.1\,M_\odot$. 

These results present conflicting mass estimates, with growing tension between higher ($\sim1.5-3M_\odot$) masses inferred from relatively bright detections (`just' a few thousand times fainter than Betelgeuse), and RV-favored masses of $0.5-1.5M_\odot$ (for low-to-modest inclination) which are also preferred by the non-detections in the FUV and X-rays. 
In this work, we present a model that would bring these apparently conflicting observational inferences into alignment: 
Betelgeuse is very bright, and the companion creates a bow shock and wake in Betelgeuse's extended chromosphere and dusty wind --- 
thus Betelgeuse's light could scatter off the wake towards the observer like a kind of diffuse `mirrorball'. 
If true, then resolved images of the shock front can be used to constrain the companion's mass, motivating future observations when $\bbud$ reappears from behind Betelgeuse in middle-to-late 2027. 

In Section \S\ref{sec:prevobs} we discuss these previous observational constraints on $\bbud$ in greater detail, highlighting especially the difficulty of extracting a mass from photometric observations. We then present a simplified model for Betelgeuse's light reflected off the companion's bow-shock in \S\ref{sec:ourpicture}, and discuss its implications in \S\ref{sec:discussion}. We summarize and conclude in \S\ref{sec:conclusion}.

\begin{table*}
    \centering
    \caption{Summary of brightness-based constraints on the companion to Betelgeuse. Magnitude constraints assume a distance of 168pc following \citet{Joyce2020}.}
    \label{tab:observational_constraints}
    \begin{tabularx}{\textwidth}{
        >{\centering\arraybackslash}p{2.5cm}
        >{\centering\arraybackslash}p{1.8cm}
        >{\centering\arraybackslash}p{1.8cm}
        >{\centering\arraybackslash}p{1.2cm}
        >{\centering\arraybackslash}p{3.0cm}
        >{\centering\arraybackslash}p{2.9cm}
        >{\centering\arraybackslash}p{1.9cm}
    }
        \toprule
        Instrument &
        Filter Name &
        Wavelength &
        FWHM &
        Constraint &
        Estimated mass ($M_\odot$) &
        Reference \\
        \midrule

        VLT/SPHERE 
        & Cnt\_Ha
        & 6449\,\AA
        & 40\,\AA
        & $1.0$--$0.75$\,mag
        & $\approx$2.6--3.1\,$M_\odot$ 
        & \citealt{Montarges2026} \\

        Gemini North `Alopeke Speckle Imager
        & EO\_466
        & 4660\,\AA
        & 440\,\AA
        & $\approx3.4$--$1.4$\,mag
        & $\approx$1.4--2$\,M_\odot$
        & \citealt{Howell2025} \\

        HST STIS 
        & FUV-MAMA E140M
        & 1144\,\AA\ -- 1710\,\AA\
        & spectrum
        & $\gtrsim 7.9$\,mag
        & $\lesssim1.5\,M_\odot$
        & \citealt{Goldberg2025} \\

        Chandra X-ray Observatory
        & HRC-I
        & 1 keV
        & --
        & $\lesssim 3.7\times10^{-4}$\,counts/s 
        & Not NS/BH
        & \citealt{OGrady2025} \\
        \bottomrule
    \end{tabularx}
\end{table*}

\section{Constraints from previous works}\label{sec:prevobs}
Here we discuss the constraints prior recent studies places on the properties of Betelgeuse as a binary system. We summarize these constraints in Table~\ref{tab:observational_constraints}.

Using photometric and radial velocity (RV) data, \citet{Goldberg2024} infer an orbital separation of 1850$\pm70\,R_\odot$, or 2.43\,$R_{\mathrm{B}}$ where $R_B$ denotes the radius of Betelgeuse, and inclination-weighted mass $m\sin i=1.17\pm0.07\,M_\odot$\footnote{The published version of \citet{Goldberg2024} has a typographical error in the abstract, which incorrectly identified $m\sin i=1.17\pm0.7\,M_\odot$ (erratum pending).}. 
From a larger baseline of RV data, \citet{MacLeod2025} infer a companion mass $m\lesssim1.25\,M_\odot$, with an RV-preferred fit of $m=0.6\,M_\odot$, and a semimajor axis of $\approx 1818\,R_\odot$, or 2.3\,$\,R_B$. 
From astrometric data, \citet{MacLeod2025} further infer that the inclination $i \approx 98^\circ\pm5^\circ \sim 90^\circ$ (nearly edge-on), and thus $\sin i \sim 1$.
\citet{MacLeod2025} also find an astrometry-only preferred mass estimate of $m\sin{i}= 2.1\pm{0.5}\,M_\odot$, but at lower confidence relative to the null hypothesis than their RV inference.
These results broadly imply a companion mass between $\approx0.5-2\,M_\odot$ and a projected orbital separation at quadrature of $\sim49-53$ mas assuming a distance of 168 pc \citep{Joyce2020}.

\citet{OGrady2025} observed the system with the Chandra X-ray observatory and report a non-detection after a 41.85\,ks exposure time. Assuming a hydrogen column density of N$_{\mathrm{H}}=6\times10^{22}$\,cm, they find an X-ray luminosity upper limit of L$_{\mathrm{X}}\lesssim10^{30}$ erg\,s$^{-1}$, ruling out an accreting compact object for the companion's intrinsic nature.\footnote{This is corroborated by \citet{Zhang2026}, who simulated a population of RSG-neutron star (NS) systems with the rapid binary population synthesis code COSMIC \citep{Breivik2020}, but failed to create a Betelgeuse+NS analog simultaneously matching the orbital properties and systemic velocity of the system.} 
Due to the wide range of observed X-lay luminosities for accreting YSOs, a constraint on YSO from X-rays was not possible. However, the authors suggest a small YSO to be more likely based on the overall trend of decreasing YSO X-ray luminosities for decreasing mass. 

\citet{Goldberg2025} observed the system with HST in far-UV (1200--1700\,\AA) with the STIS instrument, searching for a signature of the companion in emission lines with an RV offset relative to Betelgeuse's own chromospheric emission lines. 
They compare to observations of YSOs from the ULLYSES survey \citep{RomanDuval.J.2022.ULLYSES,ULLYSES2025}, under the likelihood that a stellar object of M$\sim 0.5-2$\,M$_\odot$ would, assuming coeval evolution, not have left the pre-main sequence during the $\sim$10\,Myr lifetime \citep{Joyce2020} of Betelgeuse. 
\citet{Goldberg2025} also report a non-detection, finding no excess UV flux nor any shifted emission lines at the relative velocity of the companion. The authors thus rule out an active YSO more massive than $\sim1.5$\,M$_{\odot}$, at a continuum flux limit of $\approx10^{-14}$ erg\,s$^{-1}\mathrm{cm}^2\mathrm{\AA}^{-1}$.

The first `probable' direct detection of the companion was reported by \citet{Howell2025} using the `Alopeke speckle imager on the Gemini North telescope. The authors report a luminous feature at a separation of 52\,mas and position angle of 115$^\circ$ E of N, which would be at 1.5$\sigma$ significance on its own, but in good agreement with the dynamical predictions of \citet{Goldberg2024,MacLeod2025}. 
They also include data from observations during the Great Dimming in 2020, which reveals no companion. This is expected, as the Great Dimming took place near the luminosity minimum of both Betelgeuse's LSP and FM, thus the companion was occultated behind the disc of Betelguse at that time. From a magnitude difference of $\approx$6\,mag at 4660\,\AA, \citet{Howell2025} infer a magnitude for the companion of $M_{B}\approx2.4\pm1$\,mag. Assuming coeval evolution and a system lifetime of 10\,Myr, from MIST isochrones \citep{Dotter2016,Choi2016} they infer an effective temperature of $T_\mathrm{eff}\approx$6000--10,000\,K, and a mass of $\approx1.4-2\,M_\odot$. Their preferred fit is of a 1.6$\,M_\odot$ star with $T_\mathrm{eff}\approx$7400\,K. 

\citet{Dupree2026} provided further evidence for the potential presence of a companion using both optical (HERMES/TRES) and UV (HST/STIS) spectroscopy over a long baseline. They find variations in both optical absorption and UV emission consistent with the system's LSP in period and phasing. The authors argue that these variations can be attributed to a wake within Betelgeuse's chromospheric outflows and dusty surroundings, left by the companion as it orbits. As the wake expands, it can reach the outer dust-forming region and obscure the disc of Betelgeuse. Variations in the observed ellipticity of Betelgeuse's radio image were also found to correlate with the LSP period, providing further evidence of atmospheric perturbations caused by the companion \citep{Matthews2026}. 

Most recently (at the time of the writing of this manuscript), \citet{Montarges2026} reported a 6.1$\sigma$ interferometric detection of the companion using the Very Large Telescope SPHERE-ZIMPOL instrument.
They report a separation of 52.32$\pm$0.18\,mas and position angle of 117.12$\pm$0.60$^\circ$, consistent with \citet{Howell2025}, a flux ratio of (8.24$\pm$1.04)$\times 10^{-4}$. 
Notably, they observe excess flux in the continuum H-$\alpha$ (Cnt\_Ha) filter at 6449\,\AA{} (width of 40\,\AA), 
but not in a narrow filter centered only on H-$\alpha$ at 6563.4\,\AA{} (width of 9.7\,\AA). 
Therefore they identify the emission with the continuum of the companion, rather than shock-emission which would produce an excess in excited H-$\alpha$. 
Comparing the Cnt\_Ha data to \texttt{PHOENIX} \citep{Husser2013} model spectra\footnote{Notably, the Phoenix model spectra display H-$\alpha$ absorption features rather than emission}, they infer $m\approx2.6-3.1\,M_\odot$. 
This mass range would imply the companion in a young main sequence star, rather than a YSO.

\subsection{Mass Estimates from Isochrone Fitting}

\begin{figure*}
    \centering
    \includegraphics[width=\linewidth]{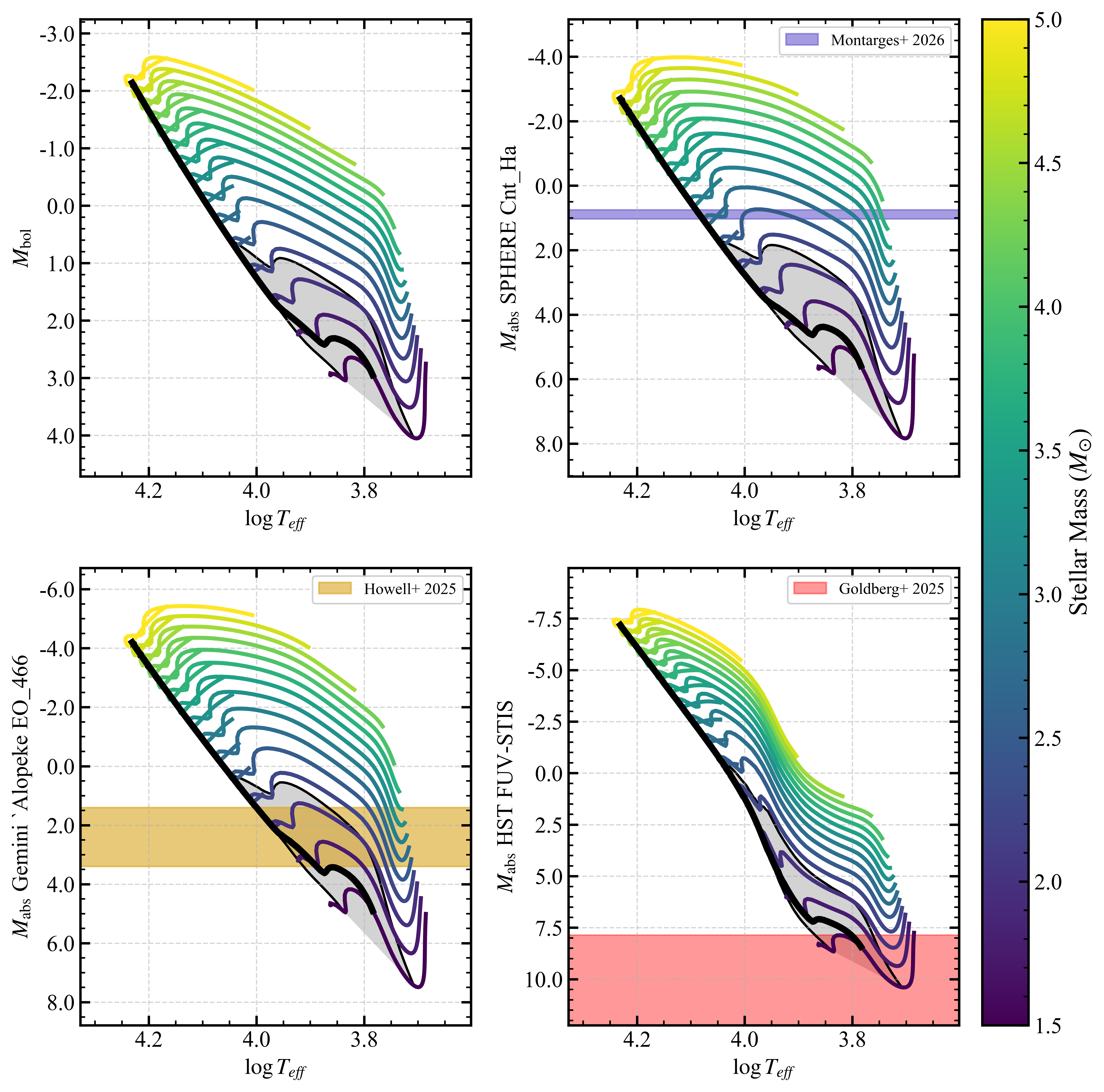}
    \caption{\textbf{Isochrone-based Mass estimates for $\bbud$:} Stellar evolution tracks colored by mass (right colorbar) and isochrones at Betelgeuse's approximate age of 10$\pm$5 Myr (10Myr: thick black curves, $\pm$5Myr: gray shaded regions) compared to the observational constraints on companion luminosity. 
    We compare directly in the observations' respective bands: 
    VLT/SPHERE detection in Cnt\_H$\alpha$ (\citet{Montarges2026}; slateblue region, upper right), 
    Gemini `Alopeke `probable' detection in EO\_466
    (\citealt{Howell2025}, goldenrod region, lower left), 
    and approximate HST FUV constraints in STIS E140M converted to FUV-MAMA (\citealt{Goldberg2025}; red region, lower right). 
    For reference, we show bolometric magnitudes at $168$pc (upper right). 
    All magnitudes are calculated using the Vega magnitude system. 
    The intersection of the isochrone-favored region (black curve + grey shading), with the observational constraints (shaded colored bars) yields the approximate range of mass estimates from each set of observations.
    }
    \label{fig:four_buddies}
\end{figure*}

The mass of $\bbud$ is not directly observed and thus is not possible to infer directly from the observations. However certain possibilities may, in theory, be excluded by constraints from non-detections (e.g., Figure 9 of \citealt{Goldberg2025}) or preferred by putative detections \citep{Howell2025, Montarges2026}. 
However, the method used by \citet{Goldberg2025}, \citet{Howell2025} and \citet{Montarges2026} to make mass predictions involved fitting to isochrones 
(or the mock stellar spectra underlying them), which is a necessary but notable weakness in all cases. As isochrone-based inferences rely on precise positioning of the star in the color-magnitude diagram, any potential misestimate or mis-attribution of the companion's brightness will be folded into and impact the companion's mass derivation.
Fitting an isochrone to a single filter, moreover, entails the unchecked assumption that the companion star's SED follows 
that of an isolated single-star track, in addition to the assumption that all of the observed flux comes directly from the companion. For main-sequence stars, this comes with the corollary that if something is brighter in redder bands, it will be even brighter in bluer bands. 

To compare the isochrone-based masses (or exclusions on such) for the companion across various observational campaigns self-consistently, we
construct our own suites of custom isochrones computed directly in the filter systems of the relevant observations using the recently released \texttt{MESA Custom Colors} \citep{CustomColors} module. We
adopt in all cases the distance of 168 pc from \citet{Joyce2020} for these calculations to place them on absolute magnitude scales. We use the Vega zero-point and corresponding SED and the Kurucz model atmospheres \cite{Kurucz1979}. Our \texttt{Custom Colors} configuration and other modeling choices, along with the grids themselves, are available on GitHub\footnote{Repository URL:\\ \texttt{https://github.com/mjoyceGR/Betelbuddy\_isochrones/}}.

We construct the grids of stellar evolutionary tracks using \texttt{MESA} version 26.05.1 \citep{Paxton2011,Paxton2013,Paxton2015,Paxton2018,Paxton2019,Jermyn2023} at solar metallicity and spanning masses from 1.5 to 5\,$M_\odot$.
The assumption of coevality between $\alpha$~Orionis~A and B underlies our modeling choices, meaning the components are taken to be the same age and born of material of the same composition. We therefore assume the companion to have solar metallicity and an age between 5 and 15\,Myr, reflecting the canonical $\sim10$\,Myr age of Betelgeuse with generous uncertainties.

We interpolate the \texttt{Custom Colors} tracks to construct filter-specific isochrones of ages 5, 10, and 15\,Myr for each of: absolute bolometric magnitude, Gemini `Alopeke  EO\_466, 
VLT/SPHERE CntH$\alpha$, and HST FUV-STIS. This produces isochrones and stellar tracks which can be compared directly to the observations, without relying on converting observations taken with custom instruments into commonly-used bands. The results are shown in Figure \ref{fig:four_buddies}. 
Additionally, we show the absolute magnitude constraints (1-$\sigma$ where applicable) at a system distance of 168pc \citep{Joyce2020}, which we calculate directly from the reported fluxes using the zero-points from SVO's filter profile service \citep{Rodrigo2012, Rodrigo2020, Rodrigo2024}.
To compare HST spectra to integrated isochrones in STIS-FUV, 
we adopt the $\approx10^{-14}$ erg\,s$^{-1}\mathrm{cm}^2\mathrm{\AA}^{-1}$ conservative 
upper limit from the spectra, and convert to magnitudes using the FUV-MAMA band zero-point. 
Observations are shown as horizontal colored bars. Isochrones are shown as black lines, with a grey shaded region indicating the $5-10$Myr range. 
The intersection of the colored bars with the isochrones yields the mass one would infer for if $T_{\rm eff}$ is unknown. The intersecting regions shown are consistent with the reported mass constraints from their respective works, summarized in the penultimate column of Table~\ref{tab:observational_constraints}. As the observing band gets bluer, the apparent mass estimate from isochrone-fitting reduces.  

\subsection{HST Constraints on Hot Main-Sequence Stars}

\begin{figure}
    \centering
    \includegraphics[width=0.95\linewidth]{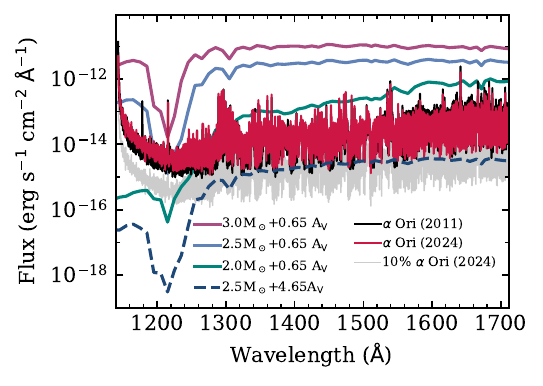}
    \caption{Far UV spectroscopy of Betelgeuse and main sequence models. HST STIS observations of Betelgeuse from 2024 \citep{Goldberg2025} are shown in red, and 2011 \citep{Carpenter2018.ASTRALBetel} in black. ATLAS9 models \citep{CastelliKurucz2003} for 3, 2.5, and 2$M_\odot$ main sequence stars are shown in violet, blue, and green, respectively, with the typical extinction for Betelgeuse of A$_\mathrm{V}$=0.65 and R$_\mathrm{V}$=4.2 applied. For the 2.5M$_\odot$ to reach $\sim10\%$ of Betelgeuse's flux (indicated in light grey), at least an additional 4 magnitudes of extinction must be applied (dark blue dashed line).}
    \label{fig:uv_plot}
\end{figure}

In the analysis of FUV HST observations conducted by \citet{Goldberg2025}, the possibility that the companion was a main-sequence (MS) star 
received little discussion due to the stellar evolutionary constraint that a star $\lesssim2M_\odot$ coeval with Betelgeuse would be near or just past its pre-MS. 
Coupled with that study's primary focus on the lack of emission features (which would have been expected from a pre-MS star or YSO), the upper limits on companion mass quoted there should not be assumed to apply to main-sequence companions. 
However, the HST noise floor does place robust upper limits on the companion's FUV continuum flux. This constraint is shown in the lower right panel of Fig.~\ref{fig:four_buddies}, where the conservative FUV upper limit of $\approx10^{-14}$ erg\,s$^{-1}\mathrm{cm}^2\mathrm{\AA}^{-1}$  is converted to a magnitude in the FUV-MAMA bandpass and then compared with isochrones in that filter. 

In Figure \ref{fig:uv_plot} we show HST far\mj{-}UV observations of Betelgeuse from 2011 \citep{Carpenter2018.ASTRALBetel} near the companion's transit, and the integrated observations from late 2024 \citep{Goldberg2025}, near the companion's quadrature. 
These spectra show bright chromospheric emission from Betelgeuse, though any evidence of continuum emission is within the noise.
We also plot main sequence model FUV spectra from ATLAS9 \citep{CastelliKurucz2003}\footnote{Accessed from \url{https://www.stsci.edu/hst/instrumentation/reference-data-for-calibration-and-tools/astronomical-catalogs/castelli-and-kurucz-atlas}} corresponding to masses of 3\,$M_\odot$ ($T_\mathrm{eff}=$11,500\,K), 2.5\,$M_\odot$ ($T_\mathrm{eff}=$10,250\,K), and 2\,$M_\odot$ ($T_\mathrm{eff}=$9000\,K). 
These models are at solar metallicity, have $\log(g)=4.0$, and have had extinction added with the \texttt{dust\_extinction}\footnote{\url{https://dust-extinction.readthedocs.io/en/stable/}} package \citep{Gordon2024} using the Milky Way extinction law of \citet{Gordon2023} 
with A$_\mathrm{V}$=0.65\,mag and R$_\mathrm{V}$=4.2 \citep{Montarges2021}. In order for a 2.5M\,$_\odot$ main sequence star to be suppressed in UV flux to \textit{approach} the level of $\sim10\%$ of Betelgeuse's UV flux, an additional $\gtrsim$4 magnitudes of extinction would be required (comparing the blue dashed line to the faint gray line). This would require nearly 100$\times$ the dust mass at fixed temperature and opacity. 
The HST FUV data are therefore indeed in tension with optical-IR luminosity-based mass estimates which prefer a companion mass of $m\gtrsim1.5\,M_\odot$, especially with the VLTI measurements which prefer $m\gtrsim2.5\,M_\odot$. 

\section{Reconciling the observations: Betelgeuse's reflection}\label{sec:ourpicture}

\begin{figure}
    \centering
    \includegraphics[width=\columnwidth]{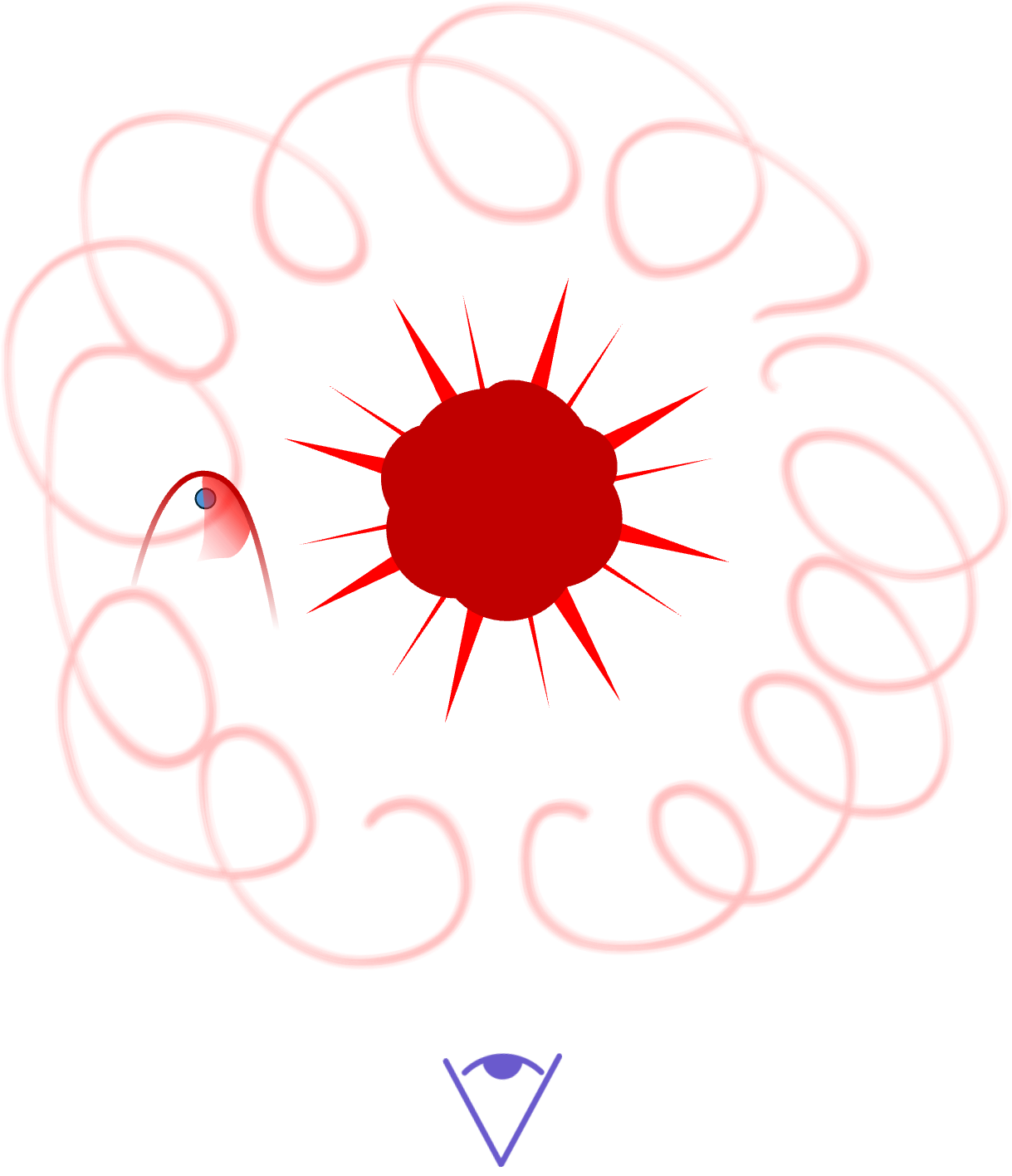}
    \caption{Schematic diagram (not to scale) of Betelgeuse (center, red blob) shining (red radial rays) on its circumstellar environment (faint curly peripheral lines), along with $\bbud$ (left, blue circle) and its bow shock/wake (left, dark red curve). For observations taken near the end of 2024, a distant Earth-based observer (bottom, purple eye) sees the Betelgeuse system configuration near $\bbud$'s receding quadrature. 
    In an environment with high-albedo dust, the companion's wake can scatter light (red patch on the right side of the wake) towards the observer.}
    \label{fig:schematic}
\end{figure}

To characterize $\bbud$ and motivate future observational campaigns, we must reconcile the following conflicting observations: the (`probable' and 6$\sigma$) detections of optical/IR-wavelength emission at or around the predicted location of the companion near receding quadrature \citep{Howell2025,Montarges2026}, 
the constraining upper limits placed by HST on the UV flux of the companion itself \citep{Carpenter2018.ASTRALBetel,Goldberg2025}, and observational evidence of periodic chromospheric and aatmospheric changes on the LSP timescale \citep{Dupree2026,Matthews2026}. 
Here, we offer a parsimonious
explanation: the emission detected by \citet{Montarges2026} and tentatively (``probably") detected by \citet{Howell2025} comes from the dense shock and surrounding wake produced as $\bbud$ travels through Betelgeuse's dusty `halo' of circumstellar material, illuminated by the bright central star. 
This is schematically illustrated in Fig.~\ref{fig:schematic}. 

\subsection{The reflective bow-shock model}
For an `object' of characteristic radius $R_{\rm c}$ at a distance $D$ from an illuminating source of luminosity $L$ (in this case, Betelgeuse), we can estimate the incident luminosity $L_\mathrm{i}$ as a function of $L$ as the ratio of the flux at $D$ to the projected surface area subtended by $R_{\rm c}$. Assuming the projected area $\sim\pi{}R_c^2$ (which we take as the definition of $R_c$), 
\begin{equation}
    \label{eq:incidentflux} 
    L_\mathrm{i} \approx L\times\frac{\pi{}R_c^2}{4\pi{}D^2}.
\end{equation}
 
If the object reflects, scatters, or reprocesses the incident light and emits it towards a distant observer, the observed luminosity $L_{\rm obs}$ from the `reflection' can be expressed as
\begin{equation}
    \label{eq:observedflux} 
    L_{\rm obs} = f_\mathrm{alb}\times{}f_\mathrm{geo}\times L_\mathrm{i} = 
    \frac{1}{4}f_\mathrm{alb}f_\mathrm{geo}\left(\frac{R_c}{D}\right)^2\times L,
\end{equation} where
$f_\mathrm{alb}$ is an `albedo' factor corresponding to the fraction of the incident light will be scattered or reflected (and thereby seen by the observer) and $f_\mathrm{geo}$ is an additional geometric factor which is a function of the location of the observer relative to the reflecting parcel and the light source. Although $f_\mathrm{geo}$ and $f_\mathrm{albedo}$ are related in nature and can be expressed as a single scaling factor, it is useful for the sake of estimation to separately consider intrinsic ($f_\mathrm{alb}$) and orientation-related ($f_\mathrm{geo}$) values. 

A purely geometric estimate of the reflectance off a perfect diffuse sphere would be \citep[][]{Russell1916}
$f_\mathrm{geo} = [\sin(\alpha) + (\pi-\alpha)\cos(\alpha)]/\pi,$
where $\alpha$ is the phase angle between the light source, the reflecting body, and the observer. 
At quadrature ($\alpha=\pi/2$), this is equal to $f_\mathrm{geo}=1/\pi\approx0.318$.\footnote{For reference, the Earth's Moon has a reasonably matte surface compared to a perfect `mirrorball,' which would be incorporated in this model into $f_\mathrm{alb}$. The Moon is also significantly brighter at its full phase due to back-scattering of light. Yet, it is still the case that $f_\mathrm{geo}\sim0.1$ in the sense that the quarter-moon appears $\approx$10\% as bright as the full moon \citep{Krisciunas1991}.}

A detailed dust radiative transfer calculation to calculate $f_\mathrm{alb}$ is beyond the scope of this work, but we note that at optical wavelengths (encompassing the observations of \citealt{Howell2025} and \citealt{Montarges2026}), absorption and scattering must be considered independently.
The ambient medium in which $\bbud$'s orbital path lives is a complex environment composed of multiphase gas and dust \citep[see, e.g.,][]{OGorman2017, Montarges2021}.
At a distance $D\sim1850\,R_\odot\approx2.5\,R_B$, this is within Betelgeuse’s extended chromosphere \citep[see discussions in][]{Goldberg2025,Dupree2026,Matthews2026}. 
Assuming a typical RSG wind \citep{Beasor2020, Beasor2023, Decin2024, Antoniadis2024} at $\sim10^{-6}M_\odot\,\mathrm{yr}^{-1}$ wind at $\sim$10-20 km/s, densities are expected to be $\sim$a few$\times10^{-13}$ in this location, and the gas-to-dust ratio is typically taken to be $\sim 400-1000$ \citep{Cannon2023}.

Dust grains within the atmospheres of RSGs are generally dominated by oxygen-rich molecules and silicates. At optical and near-IR wavelengths ($\sim0.5\mu$m, of order the grain size), 
these molecules have albedo near unity, defined in terms of opacities to absorption $\kappa_a$ versus scattering $\kappa_s$ as $\kappa_s/(\kappa_s+\kappa_a)\sim1$ \citep[e.g.][]{Tsuji1978,Hofner2007,Hofner2008,Bladh2012,Lomax2025}. 
That is to say, RSG circumstellar dust tends not to be strongly absorptive, but scatters rather efficiently. 
The exact scattering angle distribution depends on the grain lattice structure and typical orientation relative to the source; this is captured in Eq.~\ref{eq:observedflux} by deviations of $f_\mathrm{geo}$ from the perfect diffuse sphere value. 

In the Betelgeuse system specifically, in the `MOLSphere' at the location of the companion $D\approx2.5\,R_B$, the presence of alumina (Al$_2$O$_3$) is often invoked to explain the star's IR emission \citep[see, e.g.][]{Verhoelst2006,Perrin2007,Cherchneff2013}.
Absorption is known to be low for alumina, rendering radiation pressure on the grains generally insufficient to drive a stellar wind \citep[see, e.g.,][]{Hofner2016} and contributing little to the overall dust extinction, but scattering remains high, with an of albedo very nearly~$1$ assuming Mie theory \citep{Mie1908,Bauer1964}. 
We therefore, with nontrivial optical depth, we expect $f_\mathrm{alb}\approx1$, and likely not below $f_\mathrm{alb}\approx0.5$. We thus adopt $f_\mathrm{alb}=0.8$ as a reasonable-to-low fiducial value. 

We now estimate the approximate size of the bow shock region surrounding $\bbud$ to choose an appropriate value for $R_c$. In the case of a supersonic body traveling through a low-density environment, a lower bound on the characteristic radius will be the standoff distance $R_\mathrm{so}$, defined as the distance between the center of the sphere and the apex of the bow shock. For a gravitating sphere, relevant in astrophysical contexts such as this one, $R_\mathrm{so}$ depends on the object's mass, velocity, and properties of the ambient medium. 
Specifically, $R_\mathrm{so}$ is a function of the body's own radius $r$, the accretion radius 
\begin{equation}
R_A=2\,G\,m/v_{\infty}^2
\end{equation}
where $m$ and $v_\infty$ are the mass and velocity of the body, and Mach number $\mathcal{M}=v_\infty/c_s$ where $c_s$ is the ambient sound speed. More precisely, depending on the nonlinearity of the shock parameterized by $\eta$, the standoff radius is determined by \citep{Farris1994,Thun2016,Prust2024}:
\begin{equation}
\frac{R_\mathrm{so}}{r} = 
\begin{cases} 
1 + 1.1\frac{(\gamma-1)\mathcal{M}^2 + 2}{(\gamma+1)\mathcal{M}^2} \approx 1.275
& \eta \lesssim 1 \\
\eta & \eta \gtrsim 1 
\end{cases}
\end{equation}
where
\begin{equation}
\eta = \frac{1}{2} \frac{\mathcal{M}^2}{\mathcal{M}^2 - 1} \frac{R_A}{r}
\end{equation}
and 
the gas is treated as ideal with $\gamma=5/3$.

Taking the limit of high $\mathcal{M}$ and expressing $\eta$ in terms of $r$ and $R_A$, 
\begin{equation}
\label{eq:strongshockradius}
R_\mathrm{so} \approx 
\begin{cases} 
1.275\, r 
& r \gtrsim R_A/2 \\
R_A/2 
& r \lesssim R_A/2 
\end{cases}
.
\end{equation}
We note that in the limit of high $\mathcal{M}$, the regime $\eta\lesssim1$ is achieved only when $r$ exceeds $R_A/2$. Moreover, $R_\mathrm{so}$ is always of the same order-of-magnitude as the \textit{larger} of [$r$, $R_A/2$]. 

Although the radius of $\bbud$ itself is unknown, its velocity relative to Betelgeuse's wind will be approximately its orbital velocity, $v_\infty\approx43.1$\,km/s \citep[adopting orbital parameters from][]{Goldberg2024}. 
Thus, as a function of $\bbud$'s mass $m_B$, its accretion radius is $R_{A,B} \approx 205\,R_\odot\times(m_B/M_\odot)$. 
This entails a standoff radius of approximately $\approx 103\,R_\odot \times(m_B/M_\odot)$ for the companion's bow shock (or greater, for $\eta\lesssim1$), which is itself a strong lower bound on the characteristic radius $R_c$ one should adopt in Eq.~\ref{eq:incidentflux}. In fact, the spherically-averaged radius for a bow shock is larger than $R_\mathrm{so}$ by a factor of $\sim2$ \citep{Thun2016,Prust2024}.

\begin{figure*}
    \centering
    \includegraphics[width=0.95\linewidth]{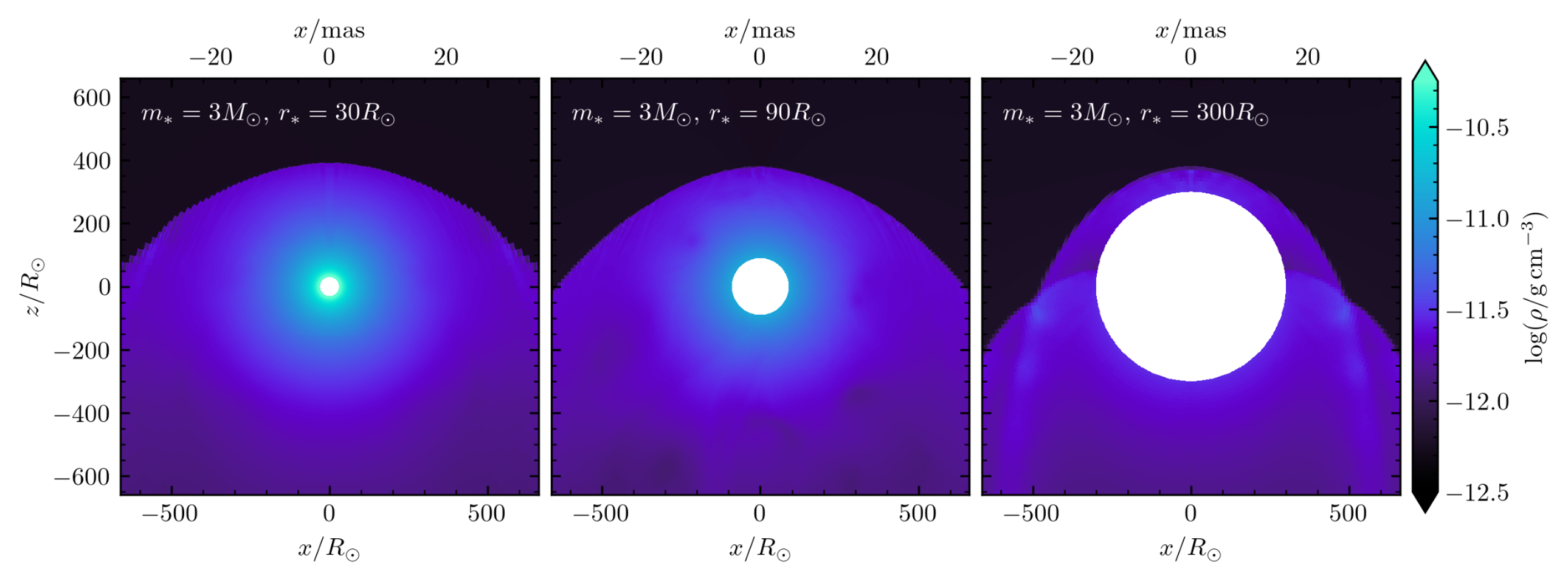}
    \includegraphics[width=0.95\linewidth]{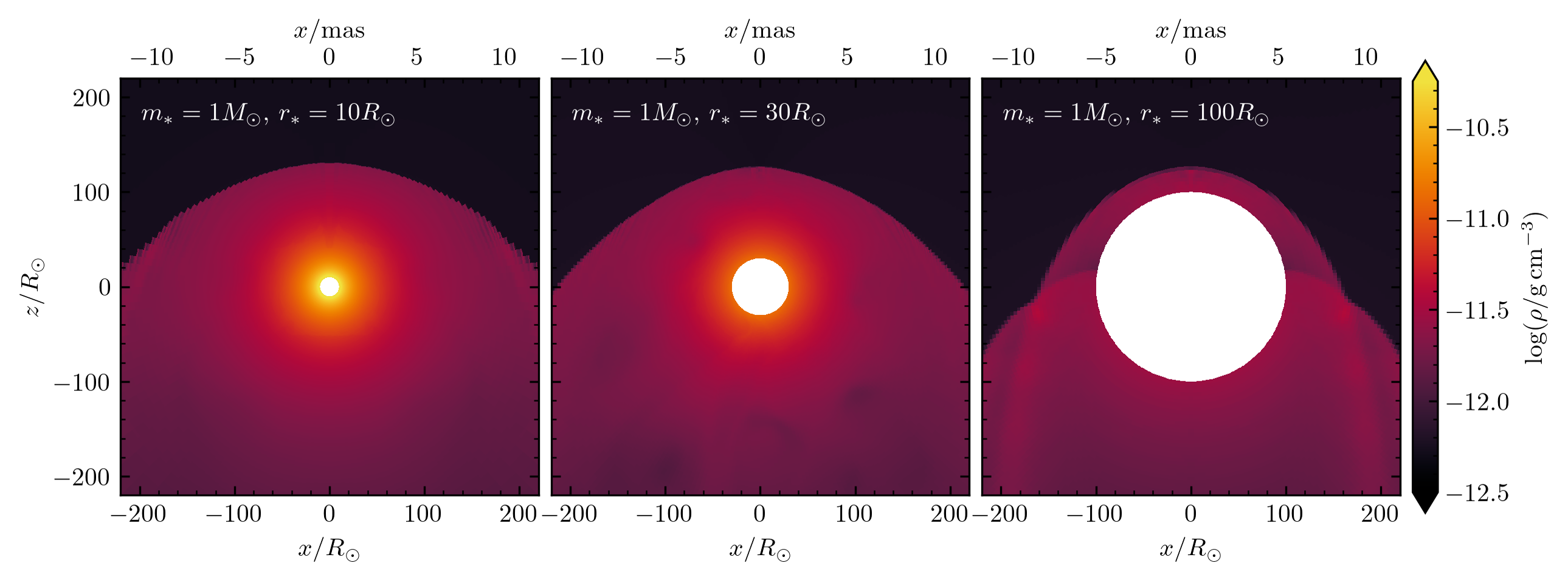}
    \caption{(Upper panel, bluer colors): \texttt{Athena++} simulations of the companion envelope assuming a companion mass of $3\,M_\odot$ (accretion radius $616\,R_\odot$), consistent with the isochrone-based mass estimate of \citet{Montarges2026}, and 3 values of the companion radius. (Lower panel, redder colors): Another set of simulations in which the companion radius and mass (and therefore the accretion radius) have been reduced by a factor of 3. 
    In all cases, the shock stand-off distance is roughly $R_A/2$.}
    \label{fig:athena}
\end{figure*}

To illustrate the shock geometry, we show \texttt{Athena++} \citep{2020ApJS..249....4S} 
hydrodynamical simulations of a spherical companion interacting with a supersonic wind in a constant density environment, adopting the wind-tunnel setup of \citet{Prust2024} in Fig.~\ref{fig:athena}. 
Here the mass and radius of the companion are varied, while the properties of the wind are held fixed assuming a static ambient density of 5.6$\times$10$^{-13}$ g\,cm$^{-3}$. 
For each value of the mass, three radius values are chosen corresponding to $\eta=10.06$, 3.35, and 1.06. As evident in the figure, all runs where $r\lesssim{}R_A/2$ nonetheless show very similar locations for the shock front for a given mass, which spreads out from $R_\mathrm{so}\approx{}R_A/2$ and widens substantially as it expands.
For $m=3\,M_\odot$ ($m=1\,M_\odot$) case, we see the shock front flaring out from the standoff radius at $\approx300\,R_\odot$ ($\approx100\,R_\odot$), approximately doubling in distance from the centroid by the time the shock passes the companion's equator.

It is also worth noting that, in a gas-dominated shock at high $\mathcal{M}$, the shock front will be $\approx4\times$ denser than the ambient medium. 
Behind the shock front, the density will continue to increase with decreasing distance from the companion \citep{Thun2016}. 
Assuming Betelgeuse’s circumstellar environment is order-of-magnitude approximated by a steady $\sim$10\,km/s wind at $10^{-6}\,M_\odot/$yr, the optical depth $\tau$ for a column of thickness $\sim100\,R_\odot$ 
and an opacity\footnote{While the dust opacity depends on the grain size and geometry (see Eq.~9 of \citealt{Hofner2016}), for radiation $\sim0.5\,\mu$m corresponding to optical light ($\sim500$\,nm), the scattering opacity per gram of alumina can be as large as $\sim10^4$\,cm$^2$/g in the optical \citep[see, e.g.][]{Hofner2008,Budaj2015}.} even of order 
$\sim$0.1 cm$^2$/g will be $\tau\sim\kappa\rho{}R_c\sim0.4$. 
Thus a factor-of-4 density increase in the vicinity of the companion could easily bring the optical depth (and thereby likelihood of at least one scattering) above unity. Even if alumina makes up only a fraction of the ambient grain distribution, an overdense region will appear over-luminous as incident light scatters off the grains concentrated near the companion's bow shock. 

In the case of the Betelgeuse system, we therefore estimate the observed luminosity seen by the observer at the location of $\bbud$'s shock front, 
relative to Betelgeuse's luminosity $L_B$, as:
\begin{equation}
\frac{L_\mathrm{obs}}{L_B} \approx 1.8\times10^{-4} \times\frac{f_\mathrm{alb}}{0.8}\times\frac{f_\mathrm{geo}}{0.3}\left(\frac{R_c}{100R_\odot}\right)^2\left(\frac{D}{1850R_\odot}\right)^{-2},\\
\end{equation}
where the scaling with orbital separation is included in case of non-negligible eccentricity \citep[see discussion by][]{Decin2025}.

This scaling is approximate, given the presence of an inhomogeneous circumstellar environment, Betelgeuse's variability, and possible tidal effects. 
That said, the dimensionless strength of the tide from components $A$ and $B$ of a binary system can be expressed as \citep{Arras2012}:
\begin{equation}
    \epsilon \equiv \frac{m_A}{m_B}\left(\frac{R_{A,B}}{D}\right)^3,
\end{equation}
where $R_{A,B}$ is the accretion radius of component B (in this case, $\bbud$). The height of the tidal bulge is $\approx\epsilon R_{A,B}$. Using the definition of the accretion radius and assuming a circular orbit, we obtain
\begin{equation}
    \epsilon = \frac{8\mu^2}{(1+\mu)^3},
\end{equation}
where $\mu=m_B/m_A$ is the mass ratio. For $\mu\sim0.05$ ($\sim$0.15), which in the $\alpha$ Ori system corresponds to a $\sim$1\,$M_\odot$ ($\sim$few-$M_\odot$) companion, $\epsilon\approx0.02$ ($\approx0.12$). We therefore do not expect a large deformation of the reflecting surface due to tides. More significant morphological differences may come from the density fluctuations in the ambient medium \citep[e.g.][]{Lim1998,Perrin2004,Kervella2009,Kervella2011,Ohnaka2009,Ohnaka2011,Montarges2016,OGorman2017,Cannon2023}. This is captured by the albedo and geometric factors, and by defining the effective $R_c$ in terms of the incident area. In that sense, the order-of-magnitude estimate is robust.

Furthermore, if the observations are indeed capturing scattered light from Betelgeuse, the optical SED will be more akin to that of Betelgeuse (with additional reddening depending on the specific properties of the molecules and dust) than that of a bright blue main-sequence star. The effects of the shock itself may additionally be visible in bluer bands, within the constraints placed by \citealt{Goldberg2025,OGrady2025}. 

The fluxes of Betelgeuse and Siwarha reported by \citet{Montarges2026} are, respectively, $(4.27 \pm 0.07)\times10^{-8}$ and $(3.44 \pm 0.44)\times10^{-11}\ \mathrm{W\,m^{-2}\,\mu m^{-1}}$. This entails a luminosity ratio of $L_{\rm obs}/L\approx 8\times10^{-4}$ where we take $L_\mathrm{obs}$ to be that of $\bbud$ and $L$ to be the luminosity of Betelgeuse. 
This difference can be reconciled by a larger effective area better-characterized by $R_c\sim200\,R_\odot$, and possibly more scattering towards the observer at quadrature due to the wake geometry. 
SPHERE/ZIMPOL can achieve a resolution of 19--20\,mas in terms of PSF full width at half maximum, which is close to the diffraction limit of the telescope itself \citep{Schmid2018}. The image of Siwarha appears unresolved in the image published by \citet{Montarges2026}, therefore we may assume $\lesssim20$\,mas as an upper limit for the angular size of the reflective side of the shock. If Betelgeuse is $\approx800\,R_\odot$, that entails an apparent radius of $R_c\lesssim380\,R_\odot$. This provides a realistic upper limit for the observed $L_\mathrm{obs}/L$ due the quadratic dependence of $L_\mathrm{obs}$ on $R_c$.

If the SED of the wake is a perfect analog of that of Betelgeuse, 
there would still be some discrepancy with the observations of \citet{Howell2025}, which place the companion at $\approx6\pm1$ magnitudes near the B-band, implying a luminosity contrast of $L_\mathrm{obs}/L\approx4\times10^{-3}$. 
However, we note that the filter used in the `Alopeke instrument on Gemini North during the detection is centered at 466\,nm with a FWHM of 44\,nm. 
Importantly, this spectral window contains the H$\beta$ line at 486.3\,nm, often excited within shocked regions \citep[e.g.][]{Raymond1979,Chevalier1980,Heng2010,Morlino_2012}. 
While \citet{Montarges2026} did not detect evidence of the companion in a narrow filter at H$\alpha$ (with a higher noise floor than their detection in the Cnt$\_\text{H}\alpha$ filter), a background of extremely bright light at red wavelengths can naturally wash out ambient H$\alpha$ emission while still allowing for observable $H\beta$ emission. Further observations following up the \citet{Howell2025} study with wider-bandpass optical instruments, narrow filters targeted at H$\beta$, and generally the comparison of filters with and without likely shock-emission lines, will be able to better constrain this scenario.

\section{Discussion and Implications}\label{sec:discussion}

We now discuss the implications of this model for constraining the companion's intrinsic properties, in the broader context of the LSP phenomenon, and its implications for follow-up observations. 

\subsection{Constraints on the companion's mass}
Depending on the radial extent of the companion, the size of the shock and corresponding wake is either a function of the companion's radius or its accretion radius $r$,
as indicated by Eq.~\ref{eq:strongshockradius} and illustrated in Fig.~\ref{fig:athena}. 
In the limit that $R_c\sim r \gtrsim R_A$, the observed luminosity is a probe of the companion's radius, independent of its mass. 
That said, a companion radius substantially in excess of $R_A$ is unlikely, as it would experience significant drag which would strip the outer layers until $r\lesssim{}R_A$ \citep[see discussion by][]{Prust2024}.
In the case that $r\lesssim{}R_A$, $R_c\appropto{}R_A$, and $L_\mathrm{obs}$ will therefore scale quadratically with the companion's mass (albeit with additional geometric complexity compared to the case of an idealized sphere). 
We can now simplify our expression by incorporating all geometric and albedo factors, along with a factor encoding the relationship between $R_c$ and $R_\mathrm{so}$, into a single scaling factor $f=f_\mathrm{geo}f_\mathrm{alb}\times(R_c/R_{\rm so})^2$, 
which is unlikely to be below $0.1$ and can exceed unity since $R_c>R_{\rm so}$ due the quadratic dependence on $R_c/R_\mathrm{so}$ and the fanning out of the shock. 
For the Betelgeuse system, folding also any variations in the incident light from eccentricity into $f$, we therefore estimate:
\begin{equation}
    L_\mathrm{obs}/L_B \approx 7\times10^{-4} \times{}f\,{}\times\left(\frac{m}{1M_\odot}\right)^2.
\end{equation}

We thus see that even a $\sim1\,M_\odot$ companion as suggested by Betelgeuse's RV modulation \citep{Goldberg2024,MacLeod2025} is fully consistent with the luminosity contrast reported by \citet{Montarges2026}.
In fact, since $R_c > R_\mathrm{so}\appropto{}\mathrm{max}\,\{r,R_A/2\}$, if $\bbud$'s mass $m\approx3\,M_\odot$ as proposed by \citet{Montarges2026} from fitting main-sequence spectral models in the ZIMPOL band, $L_\mathrm{obs}$ would be a factor of $\sim$9 greater than it would be for $m\approx1\,M_\odot$. That is, Betelgeuse's ``reflection" off the wake of a $3\,M_\odot$ gravitating object would likely be significantly \textit{brighter} 
than reported by either \citet{Montarges2026} or \citet{Howell2025}. 
This further supports the notion that 3\,$M_\odot$ is an overestimate for the intrinsic mass of the companion.

\subsection{Insight into the LSP mechanism}
If the observations of \citet{Howell2025} and \citet{Montarges2026} are indeed of Betelgeuse's light scattered by $\bbud$'s bow shock and wake as proposed in this work, this lends further credence to the picture presented by \citet{Dupree2026}: the companion leaves a trailing wake which expands to obscure Betelgeuse approximately $180^\circ$ out of phase with the companion's transit. As the wake expands to encompass cooler gas further from Betelgeuse and reaches the outer dust shell (\citealt{Kervella2016}; see also \citealt{Perrin2007,Kervella2011}), it can obscure the central star. This, roughly, would be the phase of the LSP minimum. 

Additionally, detection of light at the location of the companion's wake may explain the tension in the slight misalignment of the orbital axis inferred from imaging \citep{Howell2025,Montarges2026} compared to the inferred rotation axis \citet{Kervella2018}\footnote{We also note that this tension relies on 2 important assumptions: 1) that the rotation axis measured by \citet{Kervella2018} is robust to the point that the star's large-scale convection cannot obscure the inferred rotation axis by more than the statistical error \citep[see][for an argument against such an assumption]{Ma2024}, and 2) that tidal interaction with the companion is responsible for Betelgeuse's rotation, rather than a recent merger \citep{Wheeler2017,Chatzopoulos2020, Shiber2024, Wheeler2023}, as discussed in detail by \citet{MacLeod2025}.}. More specifically, given the sizable $R_c$ relative to Betelgeuse's radius, 
emission there rather than precisely centered on the companion could appear to be a misalignment between the orbital plane and rotation axis even at quadrature (when the apparent position is less affected by inclination relative to the observer). 

One additional piece of evidence pointing to binarity in other LSP systems is the presence of secondary eclipses in the IR lightcurves of a subset of LSP systems \citep{Soszynski2014,Soszynski-2021}. This is often interpreted as an orbiting dusty cloud which contributes to the observed IR flux when on the limb, but obscures the central star and causes a dip in the IR at transit. An IR-bright dusty cloud going behind the stellar disc at occultation causes a smaller, second dip in the luminosity. In that picture, however, if the companion is embedded within the dust cloud and carries it as it orbits, the LSP optical minimum and deeper IR minimum should occur during the companion's transit. 
In the IR, primary minima occur roughly coincident with the optical minima \citep{Soszynski-2021}. 
Yet as \citet{Goldberg2024} show, in a majority of systems including Betelgeuse, the phase between the lightcurve minimum and the companion's transit is offset by half an orbit. 
This means the primary IR minima tend to occur when the companions are approximately \textit{behind} their central stars, and the shallower secondary minima occur when the companions are transiting the disk. This is difficult to reconcile with a dust cloud being carried along by the companion. 

This \textit{could} be naturally explained if other LSP systems similarly exhibit reflection off the shock front and wake preferentially at long wavelengths. 
In that case, the shocked halo surrounding the companion is illuminated by the bright (red) central star, reprocessing the light towards the observer. 
That additional reprocessed flux will be reduced when the companion is in transit (at optical maximum) due to both reduced $f_\mathrm{geo}$ and the alignment of the two IR sources --- potentially observable as a secondary IR dip.  
Typically, when observed, these secondary eclipses are on the order of $\sim0.01-0.05$\,mag, or $1-5\%$. 
This requires a substantial $R_c/D$ even for $f\sim1$, 
but depending on the mass ratio, scattering properties, wind and shock geometry, and binary separation, a few \% variation in flux is not un-achievable, and some of the additional lightcurve modulation at different phases could in principle be a result of the phase-dependence of $f_\mathrm{geo}$. 
The primary dimming then occurs half an orbit later,
caused, e.g., by the wake expanding to reach dust-forming temperatures as discussed by \citet{MacLeod2025,Dupree2026}, or by dynamical effects on the dust shell \citep{Decin2025}.  

We note also that such substantial secondary IR dips are not ubiquitous. In Betelgeuse, secondary IR eclipse behavior is at least not obviously identified \citep{Taniguchi2022,Jadlovsky2023}. However, it is most likely that the high luminosity and factor-of-2 ($\gtrsim1$ mag) variability would render a secondary eclipse at the few-$\%$ level nearly impossible to detect. 
Betelgeuse's LSP amplitude is also much too large for reflection off companion to be fully responsible for the (primary) LSP mininmum, unless $R_c$ approaches or exceeds the radius of Betelgeuse itself. More work is therefore needed to determine the extent to which companion-wake illumination explains the behavior of the subset of LSP systems exhibiting detectable secondary IR dips at optical LSP maximum. 
Nonetheless, the apparent LSP phase paradox, that the deep LSP minimum comes at the companion's occultation rather than its transit, accords nicely with the reflective bow-shock model. 

\subsection{Implications for future observations}\label{sec:observations} 
Given the possible presence and importance of reflected light from Betelgeuse near the companion, we suggest future targeted observations aimed at both discerning the luminosity \textit{and original source} of any light source near the companion's predicted position. 
We also emphasize the importance of non-detections in constraining scenarios, e.g., if blue emission can be difficult to hide. 
A $T_\mathrm{eff}\sim$11,000–12,000\,K star (as a 3M\,$_{\odot}$ main-sequence star would be) will have a blue optical SED unless the local extinction surrounding just the companion is both very large (with very low albedo dust) and \textit{strongly} wavelength-dependent, which we at present have no reason to assume. 

Likewise, in order to discern shock emission from companion emission in the \citet{Howell2025} observations, we suggest observing with a narrow H$\beta$ filter, ideally centered on H$\beta$ with a width of $\sim\pm100$\,km/s, as well as a wider blue filter. 

Additionally, now that the apparent location is more tightly constrained, it will be easier to conduct targeted observations with interferometry, speckle imaging, and potentially adaptive optics. Further resolved images at different orbital phases will provide further constraints on the companion's orbital inclination and eccentricity. 
Similarly, we may learn more from spatially-resolved FUV spectroscopic observations simultaneously taken near the companion's location in the chromosphere and near Betelgeuse's disc, as spectroscopic differences will be easier to distinguish in narrower slits than the 0.2"$\times$0.2" slits used by \citet{Goldberg2025}. 

Another interesting possibility for future observations would be spatially-resolved spectropolarimetry. 
If the observed emission is indeed scattered light off of circumstellar grains, it should be linearly polarized with a centro-symmetric/tangential polarization angle relative to Betelgeuse. 
Moreover, spectra should exhibit a wavelength-dependent degree of polarization determined by grain size and composition, and potentially be spatially extended along the wake.
In Fig.~B.1 of \citet{Montarges2026}, the degree of linear polarization observed with VLT/SPHERE-ZIMPOL is shown, and it is indeed higher near the location of the presumed companion image. Decomposing this further into Stokes Q and U parameters would be of great interest in constraining the wake geometry. 

Turning to the time-domain, a wake should also trail the instantaneous orbital position, and may exhibit a phase-dependent centroid offset. Therefore, changes in morphology within observations at different separations, beyond just the maximum separation, will be informative for constraining the shock geometry in addition to the orbital parameters.

\section{Conclusions}\label{sec:conclusion}

In this work we deconstruct and resolve the factor-of-2 tension between different mass estimates for Betelgeuse's companion from the spatially-resolved observations of \citet{Howell2025} and \citet{Montarges2026}, the non-detections of \citet{Goldberg2025} and \citet{OGrady2025}, and RV-based analyses \citet{Goldberg2024,MacLeod2025}. 
We argue that, in optical bands, the observed emission is likely to originate from reflected light from Betelgeuse rather than continuum emission from the companion. 

Thus, although ample remains to be understood, we demonstrate that an intrinsically faint companion is nonetheless consistent with the bright near-optical emission seen in imaging by \citet{Montarges2026} and `probably' by \citet{Howell2025}.
Such a scenario is moreover consistent with the lower end ($\lesssim1.5\,M_\odot$) of mass estimates from the low FUV continuum flux \citep{Carpenter2018.ASTRALBetel, Goldberg2025} and preferred by RV observations \citep{Goldberg2024,MacLeod2025} assuming relatively low inclination angles (as reported for the rotation axis; see e.g. \citealt{Dupree1987,Gilliland1996,Uitenbroek1998,Lobel-2001,Kervella2009,Kervella2018}) which may align with the companion's orbital plane).  
This is also in good conceptual agreement with the expanding wake detected by \citet{Dupree2026}, other periodic circumstellar behavior \citep{Matthews2026}, and more generally the shock properties of a supersonic gravitating object of $\sim1\,M_\odot$ at $\bbud$'s orbital separation in the presence of dust characteristic of RSG atmospheres within a few stellar radii.

Motivated by \citet{Montarges2026}'s demonstrated ability to observe emission at the companion's location, we discuss potential observational paths forward in \S\ref{sec:observations} which may distinguish light reflected off the shock front and wake (which may entail the inability to easily observe the companion's continuum directly) from the companion's intrinsic emission.
We predict high linear polarization for this material; thus spectropolarimetric observations at various orbital configurations may aid in disentangling the presence, and if so, geometry of Betelgeuse's wake through the complex circumstellar halo.  
We also suggest spatially-resolved observations at multiple phases in addition $\bbud$'s maximum elongation at approaching and receding quadrature.
Therefore, we eagerly await the next few years of observations as the companion returns from behind the disc of Betelgeuse so that we can validate, refine, or refute this model, and further investigate the shockingly bright, new-found addition to the Betelgeuse family. 


\begin{acknowledgements}
    The authors would like to thank Stephen Justham for pointing out the possibility of H-beta contributing to the \citet{Howell2025} observations, Andrea Dupree and Morgan MacLeod for valuable discussions, and Niall Miller for incorporating the relevant bandpasses into the MESA Custom Colors module.
    JAG additionally thanks Zarina Dhillon for inspiration, motivation, and illuminating conversations, the TARDIS group at MSU for discussions of radiative transfer (especially Josh Shields, Andrew Fullard, and Connor McClellan), and Iman Behbehani and Thavisha Dharmawardena for many discussions of Betelgeuse's dust. AOG thanks Adiv Paradise for helpful conversations. 

    This research has made extensive use of the Astrophysics Data System (ADS), funded by NASA under Cooperative Agreement 80NSSC21M00561.

    This research has made use of the Spanish Virtual Observatory (https://svo.cab.inta-csic.es) project funded by MCIN/AEI/10.13039/501100011033 through grant PID2023-146210NB-I00. 

    This research was supported at Michigan State University by a NASA Astrophysics Theory Program Grant 23-ATP23-0070. 
    
    The technical support and advanced computing resources from University of Hawaii Information Technology Services – Research Cyberinfrastructure, funded in part by the National Science Foundation CC* awards \#2201428 and \#2232862 are gratefully acknowledged.

    This research was supported by the `SeismoLab' KKP-137523 \'Elvonal grant of the Hungarian Research, Development and Innovation Office (NKFIH) and by the LP2025-14/2025 Lendület grant of the Hungarian Academy of Sciences.
    The Flatiron Institute is supported by the Simons Foundation.
    
\end{acknowledgements}

\textbf{Software:}
This work made use of the following software packages: \texttt{python} \citep{python}, including \texttt{numpy} \citep{numpy}, \texttt{scipy} \citep{2020SciPy-NMeth,scipy_20615351}, \texttt{matplotlib} \citep{Hunter2007}, \texttt{CMasher}, \citep{2020JOSS....5.2004V,CMasher_14186007}, Modules for Experiments in Stellar Astrophysics \citep[MESA][]{Paxton2011, Paxton2013, Paxton2015, Paxton2018, Paxton2019, Jermyn2023}, and \texttt{Athena++} \citep{Stone2020,Athena++_11660592}. We make extensive use of the new \texttt{MESA Colors} module \citep{CustomColors}.

The MESA EOS is a blend of the OPAL \citep{Rogers2002}, SCVH \citep{Saumon1995}, FreeEOS \citep{Irwin2004}, HELM \citep{Timmes2000}, PC \citep{Potekhin2010}, and Skye \citep{Jermyn2021} EOSes. Radiative opacities are primarily from OPAL \citep{Iglesias1993, Iglesias1996}, with low-temperature data from \citet{Ferguson2005} and the high-temperature, Compton-scattering dominated regime by \citet{Poutanen2017}. Electron conduction opacities are from \citet{Cassisi2007} and \citet{Blouin2020}. Nuclear reaction rates are from JINA REACLIB \citep{Cyburt2010}, NACRE \citep{Angulo1999} and additional tabulated weak reaction rates \citet{Fuller1985, Oda1994, Langanke2000}. Screening is included via the prescription of \citet{Chugunov2007}. Thermal neutrino loss rates are from \citet{Itoh1996}. Roche lobe radii in binary systems are computed using the fit of \citet{Eggleton1983}. Mass transfer rates in Roche lobe overflowing binary systems are determined following the prescription of \citet{Ritter1988}. 

Software citation information aggregated using \texttt{\href{https://www.tomwagg.com/software-citation-station/}{The Software Citation Station}} \citep{software-citation-station-paper,software-citation-station-zenodo}.

\bibliography{buddybib}{}
\bibliographystyle{aasjournalv7.1}

\end{document}